\documentclass[submission,copyright,creativecommons]{eptcs}
\providecommand{\event}{LINEARITY'16} % Name of the event you are submitting to
\usepackage{breakurl}             % Not needed if you use pdflatex only.

\usepackage{graphicx}
\usepackage{amsmath}
\usepackage{amssymb}
\usepackage{amsthm}
\usepackage{mathtools}
\usepackage{stmaryrd}
\usepackage{todonotes}

\usepackage{caption}
\usepackage{tabularx}
\usepackage{booktabs}

\title{Krivine Machine and Taylor Expansion in a Non-uniform Setting}
\author{Antoine Allioux
\institute{Institut de Recherche en Informatique Fondamentale\\ Paris, France}
\email{antoine.allioux@gmail.com}}

\def\titlerunning{Krivine Machine and Taylor Expansion in a Non-uniform Setting}
\def\authorrunning{A. Allioux}

\newtheorem{theorem}{Theorem}[section]

\theoremstyle{definition}
\newtheorem{definition}[theorem]{Definition}

\begin{document}
\maketitle

\begin{abstract}
The Krivine machine is an abstract machine implementing the linear head reduction of $\lambda$-calculus. Ehrhard and Regnier gave a resource sensitive version returning the annotated form of a $\lambda$-term accounting for the resources used by the linear head reduction. These annotations take the form of terms in the resource $\lambda$-calculus.

We generalize this resource-driven Krivine machine to the case of the algebraic $\lambda$-calculus. The latter is an extension of the pure $\lambda$-calculus allowing for the linear combination of $\lambda$-terms with coefficients taken from a semiring. Our machine associates a $\lambda$-term $M$ and a resource annotation $t$ with a scalar $\alpha$ in the semiring describing some quantitative properties of the linear head reduction of $M$.

In the particular case of non-negative real numbers and of algebraic terms $M$ representing probability distributions, the coefficient $\alpha$ gives the probability that the linear head reduction actually uses exactly the resources annotated by $t$. In the general case, we prove that the coefficient $\alpha$ can be recovered from the coefficient of $t$ in the Taylor expansion of $M$ and from the normal form of $t$.
\end{abstract}

\section{Introduction}

The Krivine machine is an abstract machine implementing the linear head reduction~\cite{DR99} on the pure $\lambda$-calculus. Ehrhard and Regnier gave a resource sensitive version~\cite{BKT} returning the annotated form of a $\lambda$-term accounting for the resources used by the linear head reduction. These annotations take the form of terms in the resource $\lambda$-calculus. As an example, the ordinary term $((\lambda x.(x)x)\lambda x.x)c_0$ which reduces to the constant $c_0$ is annotated by the following resource term $\langle \langle \lambda x.\langle x \rangle x^1 \rangle (\lambda x.x)^2 \rangle c_0^1$. This resource term informs us that $\lambda x.x$ is used twice during the reduction and $x$ and $c_0$ are used once.

We generalize this resource-driven Krivine machine to the case of the algebraic $\lambda$-calculus\footnote{This machine has been implemented and is available online at \url{http://allioux.iiens.net/taylor/}.}. The latter is an extension of the pure $\lambda$-calculus allowing for the linear combination of $\lambda$-terms with coefficients taken from a semiring. Some properties enjoyed by the ordinary $\lambda$-calculus do not hold anymore in the case of the algebraic $\lambda$-calculus and some results become nontrivial. Our machine associates a $\lambda$-term $M$ and a resource annotation $t$ with a scalar $\alpha$ in the semiring describing some quantitative properties of the linear head reduction of $M$. We will only consider terms reducing to a multiple of a constant for the sake of convenience.

In the particular case of non-negative real numbers and of terms $M$ representing probability distributions, the coefficient $\alpha$ gives the probability that the linear head reduction actually uses exactly the resources annotated by $t$. In the general case, we prove that the coefficient $\alpha$ can be recovered from the coefficient of $t$ in the Taylor expansion of $M$ and from the normal form of $t$. A more detailed report concerning this work can be found at \url{http://allioux.iiens.net/taylor/report.pdf}.

\section{Algebraic lambda calculus}

The algebraic $\lambda$-calculus is an extension of the pure $\lambda$-calculus allowing for the linear combination of $\lambda$-terms. More precisely, we endow it with a structure of left $\mathbb{S}$-module where $\mathbb{S}$ is a semiring. We shall follow the presentation of the algebraic $\lambda$-calculus given in~\cite{VAU09}.

\subsection{Grammar}

Let $x$ be a variable in $\mathcal{V}$, the set of variables, and let $\alpha$ be a scalar in $\mathbb{S}$. The grammar of the algebraic $\lambda$-calculus is the following:
\begin{equation}
  \Lambda_{\mathbb{S}} : M,N \Coloneqq x \ | \ \lambda x.M \ | \ (M)N \ | \ \alpha M \ | \ M+N \ | \ \mathbf{0}
\end{equation}

We denote $\equiv_{\text{alg}}$ the equivalence relation described in Table~\ref{tab:algebraic-rules} making $\Lambda_{\mathbb{S}}$ into a left $\mathbb{S}$-module and providing linear properties to terms. We consider the terms of the quotient set $\Lambda_{\mathbb{S}} / \equiv_{\text{alg}}$ up to $\alpha$-conversion and we call them \emph{algebraic terms}. We define free variables and $\alpha$-conversion as in~\cite{VAU09}.

  \begin{table*}
  \begin{tabularx}{\linewidth}{X}
    \toprule
    \hfill \textit{Algebraic equalities of the $\mathbb{S}$-module} \hfill \null\\
    \midrule
    \hfill $M+\mathbf{0} \equiv_{\text{alg}} M$ \hfill $(M+N)+P \equiv_{\text{alg}} M+(N+P)$ \hfill $M+N \equiv_{\text{alg}} N+M$ \hfill \null\\
    \hfill $\alpha(M+N) \equiv_{\text{alg}} \alpha M + \alpha N$ \hfill $\alpha M + \beta M \equiv_{\text{alg}} (\alpha+\beta)M$ \hfill $\alpha(\beta M) \equiv_{\text{alg}} (\alpha \beta)M$ \hfill \null\\
    \hfill $1M \equiv_{\text{alg}} M$ \hfill $\mathbf{0}M \equiv_{\text{alg}} \mathbf{0}$ \hfill $\alpha \mathbf{0} \equiv_{\text{alg}} \mathbf{0}$ \hfill \null\\
    \midrule
    \hfill \textit{Linear properties} \hfill \null\\ 
    \midrule
    \hfill $\lambda x.(M+N) \equiv_{\text{alg}} \lambda x.M + \lambda x.N$ \hfill $\lambda x.(\alpha M) \equiv_{\text{alg}} \alpha (\lambda x.M)$ \hfill $\lambda x.\mathbf{0} \equiv_{\text{alg}} \mathbf{0}$ \hfill \null\\
    \hfill $(\mathbf{0})M \equiv_{\text{alg}} \mathbf{0}$ \hfill $(\alpha M)N \equiv_{\text{alg}} \alpha(M)N$ \hfill $(M+N)P \equiv_{\text{alg}} (M)P+(N)P$ \hfill \null\\
    \bottomrule
  \end{tabularx}
  \caption{Algebraic equalities of the algebraic $\lambda$-calculus}
  \label{tab:algebraic-rules}
  \end{table*}
  
  \subsection{Algebraic states}
  
  The behaviour of the Krivine machine is defined on some structures we call \emph{states} with which we can associate a unique algebraic term rather than on algebraic terms directly. A state is a snapshot of the abstract machine at a given time and represents the dissection of a \textit{unique} $\lambda$-term.

  \begin{description}
    \item [Algebraic environment] An algebraic environment is a finite partial function $E$ mapping variables to closures. We introduce the notation $E_{x \mapsto \Gamma}$ to refer to the environment which behaves like $E$ for variables other than $x$ and which maps $x$ to $\Gamma$.
    \item [Algebraic closure] An algebraic closure $\Gamma$ is a pair $(M, E)$ composed of an algebraic term $M \in \Lambda_{\mathbb{S}}$ and of an environment $E$ such that $\mathrm{FV}(M) \subseteq \mathrm{Dom}(E)$ where $\text{FV}(M)$ denotes the free variables of $M$ and $\mathrm{Dom}(E)$ denotes the domain of $E$.
    \item [Algebraic state] An algebraic state is a nonempty stack of closures. We choose to denote states as triples $(M, E, \Pi)$ where $(M, E)$ is the first closure of the stack and $\Pi$ is the stack of the remaining closures. Indeed, our Krivine machine implementing the linear head reduction, we reduce according to the structure of the first closure and we give it a special status. We refer to the set of the algebraic states by $\mathcal{S}(\Lambda_{\mathbb{S}})$.
  \end{description}

  The intuition behind algebraic states can be made explicit by defining the function $\text{T}:\mathcal{S}(\Lambda_{\mathbb{S}}) \rightarrow \Lambda_{\mathbb{S}}$ which given an algebraic state returns its unique associated algebraic term.

  Given any algebraic closure $(M,E)$ and any stack of algebraic closures $\Gamma_1, \dots, \Gamma_n$ with $n \geq 0$, we first define T on closures and then extend it to states as follows:
  \begin{align*}
    \label{eq:T-def}
    \text{T}(M,E) &= M[\text{T}(E(x))/x]_{x \in \mathrm{Dom}(E)}\\
    \text{T}(M, E, (\Gamma_1, \dots, \Gamma_n)) &= (\dots(\text{T}(M,E))\text{T}(\Gamma_1)\dots)\text{T}(\Gamma_n)
  \end{align*}
  
%  It is possible to recover this $\lambda$-term and we define a function $\mathrm{T}:\mathcal{S}(\Lambda_{\mathbb{S}}) \rightarrow \Lambda_{\mathbb{S}}$ doing precisely that.
%  
%  T is defined on closures and extended to states as follows:
%  \begin{align}
%    \label{eq:T-def}
%    \mathrm{T}(M,E) &= M[\mathrm{T}(E(x))/x]_{x \in \mathrm{Dom}(E)}\\
%    \mathrm{T}(M,E,(\Gamma_1, \dots, \Gamma_n)) &= (\dots(\mathrm{T}(M,E))\mathrm{T}(\Gamma_1)\dots)\mathrm{T}(\Gamma_n)
%  \end{align}
  
  \subsection{Krivine machine}
  
  In this particular section the semiring $\mathbb{S}$ is complete --- its sum is infinitary.
  
  We give a description of the Krivine machine as the limit $K$ of the sequence $(K_n)_{n \in \mathbb{N}}$ defined by induction on $(n,M)$ lexicographically ordered where $n$ is a non-negative integer and $M$ is an algebraic term. The induction on $n$ turns the reduction of $M$ into a finite process even for non-normalizing terms. We also enrich the grammar of the algebraic $\lambda$-calculus with the constant $c_0$ as we restrict our study to closed terms reducing to this constant.

  \begin{itemize}
    \item $K_0(M,E, \Pi)=\mathbf{0}$,
    \item $K_{n+1}(c_0, E, \emptyset)=c_0$,
    \item $K_{n+1}(x, E, \Pi)=K_{n}(E(x), \Pi)$ if $x \in \mathrm{Dom}(E)$,
    \item $K_{n+1}(\lambda x.M, E, \Gamma::\Pi)=K_{n}(M, E_{x \mapsto \Gamma}, \Pi)$ assuming $x \notin \mathrm{Dom}(E)$,
    \item $K_{n+1}((M)N, E, \Pi)=K_{n}(M, E, (N,E)::\Pi)$.
  \end{itemize}
  
  These rules, excluding the first two ones, are the ones of the original Krivine machine. As the algebraic $\lambda$-calculus is just an extension of the ordinary $\lambda$-calculus, it suffices to add the two following rules to the description of the Krivine machine to handle it:
  \begin{itemize}
    \item $K_{n+1}(\alpha M, E, \Pi) = \alpha K_{n+1}(M, E, \Pi)$,
    \item $K_{n+1}(M+N, E, \Pi)=K_{n+1}(M, E, \Pi) + K_{n+1}(N, E, \Pi)$.
  \end{itemize}
  
  Finally we set $K=\lim_{n \to \infty} K_n$.

\section{Resource lambda calculus}

We recall the syntax and the reduction of the resource $\lambda$-calculus which has been defined in~\cite{EHR08}. Indeed, we will define the Taylor expansion of an algebraic term in terms of a sum of resource terms. 

\subsection{Grammar}

  The resource $\lambda$-calculus shares its syntax with the ordinary $\lambda$-calculus with the exception that the application takes multisets of terms as argument. We use the multiplicative notation to denote multisets so the multiplicative unit $1$ is the empty multiset. For example $s^2t$ is the multiset formed of two occurrences of $s$ and one occurrence of $t$. Multisets are commutative. The multiset union of $S$ and $T$ is denoted $ST$. The multiplicity of an element $t$ in a multiset $T$ is given by $T(t)$. The support of $T$, denoted $\text{supp}(T)$, is the set of elements of $T$ whose multiplicity is nonzero.
  
  %The cardinal of a multiset $T$, denoted $|T|$ is defined to be $\sum_{t \in \text{supp}(T)}T(t)$.

  The grammar of \emph{simple terms} in the resource $\lambda$-calculus is:
  \begin{equation}
    \Delta : s,t,u \Coloneqq x \ | \ \lambda x.t \ | \ \langle t \rangle S
  \end{equation}
  
  where $x,y,\dots \in \mathcal{V}$, the set of variables and where $S$ is a finite multiset of simple terms. We denote the set of simple terms $\Delta$ and we refer to its elements using lower case letters $s,t,\dots$. The set of finite multisets of simple terms is denoted $\Delta^!$. We call its elements simple \emph{poly}-terms and we refer to them using upper case letters $S,T,\dots$. When a term $t$ can either be a simple term or a simple poly-term we say it is in $\Delta^{(!)}=\Delta \cup \Delta^!$.
  
  When denoting an application, we use the Krivine notation which we recall: for any simple term $t$ and any simple poly-terms $S_1, \dots, S_n$, we denote the application $\langle \dots \langle t \rangle S_1 \dots \rangle S_n$ by the simplified form $\langle t \rangle S_1 \dots S_n$.
  
  The module $\mathbb{S} \langle \Delta^{(!)} \rangle$ is the set of linear combinations of simple (poly-)terms with coefficients in $\mathbb{S}$. We call its elements (poly-)terms in opposition to \textit{simple} (poly-)terms which are not part of a linear combination. These combinations can not be expressed in the syntax of the resource $\lambda$-calculus contrarily to the algebraic $\lambda$-calculus. We refer to (poly-)terms using the letters $\mathcal{S}, \mathcal{T}, \dots$. We denote $\mathcal{S}_s$ the coefficient of the (poly-)term $s$ in $\mathcal{S}$. Finally we extend the grammar of the resource $\lambda$-calculus to all (poly-)terms by multilinearity so that: $\lambda x.(t+u) = \lambda x.t + \lambda x.u$, $\langle s+t \rangle T = \langle s \rangle T + \langle t \rangle T$ and $(s+t)T=sT + tT$.
  
\subsection{Linear substitution and reduction}

The reduction of the resource $\lambda$-calculus is based on a particular notion of linear substitution. What distinguishes linear substitutions from classical substitutions is that, in the former case, substituted terms have to be used once and only once whereas this restriction does not apply in the latter case.
  
  For a variable $x$ and a term $s$, we define deg$_x(s)$ to be the number of occurrences of $x$ in $s$ and we call it the degree in $x$ of $s$.
  
Let $\hat{s}$ be the resource term obtained from a resource term $s$ by renaming its different occurrences of $x$ to $x_1, \dots, x_n$ with $n=\mathrm{deg}_x(s)$. $\hat{s}$ is such that for all $i \in \llbracket 1,n \rrbracket$, deg$_{x_i}(\hat{s})=1$ and $s=\hat{s}[x/x_1,\dots,x_n]$.
 
Let $s$ be a simple term and let $t_1 \dots t_n$ be any poly-term with $n$ being a non-negative integer, the linear substitution of $s$ by $t_1 \dots t_n$ is defined as follows: 
\begin{equation}
  \partial_x(s, t_1 \dots t_n) = 
  \begin{cases}
    \displaystyle \sum_{f \in \mathfrak{S}^n} \hat{s}[t_{f(1)}/x_1,\dots,t_{f(n)}/x_n]  & \text{ if deg}_x(s) = n\\
    \mathbf{0} \in \mathbb{S} \langle \Delta \rangle & \text{ if deg}_x(s) \neq n 
  \end{cases}
\end{equation}
with $\mathfrak{S}^n$ being the group of permutations on the set $\{1,\dots,n\}$. This construction can be extended to simple (poly-)terms. 

We extend this notation to the linear substitution of several variables.
For all poly-terms $T_1, \dots, T_n$ with $n$ being a non-negative integer, 

\begin{equation}
  \partial_{x_1,\dots,x_n}(s,T_1,\dots,T_n) = \partial_{x_n}(\dots \partial_{x_1}(s,T_1), \dots, T_n)
\end{equation}

This substitution does not depend on the order of the iterated substitutions as the variables $x_1,\dots,x_2$ are pairwise distincts.

We derive the $\beta$-reduction relation for the resource $\lambda$-calculus from this linear substitution. A redex in the resource $\lambda$-calculus is of the form $\displaystyle \langle \lambda x.s \rangle T$ and reduces as follows: $\langle \lambda x.s \rangle T \rightarrow_\beta \partial_x(s,T)$.

%\begin{equation*}
%  \langle \lambda x.s \rangle T \rightarrow_\beta \partial_x(s,T)  
%\end{equation*}

We extend this relation to $\mathbb{S} \langle \Delta^{(!)} \rangle \times \mathbb{S} \langle \Delta^{(!)} \rangle$ by defining it as being the least relation closed under the following rules, assuming $s \rightarrow_{\beta} \mathcal{S}$ with $s \in \Delta$ and $\mathcal{S} \in \mathbb{S} \langle \Delta \rangle$:\\
$\hfill \langle s \rangle T \rightarrow_{\beta} \langle \mathcal{S} \rangle T \hfill \langle u \rangle sT \rightarrow_{\beta} \langle u \rangle \mathcal{S}T \hfill \lambda x.s \rightarrow_{\beta} \lambda x.\mathcal{S} \hfill s+u \rightarrow_{\beta} \mathcal{S}+u \hfill \null$ 

This relation is confluent and strongly normalizing for $\mathbb{S}=\mathbb{N}$ as proved in~\cite{EHR03} and we derive NF, the unique normalization map $\mathbb{N} \langle \Delta^{(!)} \rangle \rightarrow \mathbb{N} \langle \Delta_0^{(!)} \rangle$, where $\Delta_0$ stands for the set of normal simple terms.

\subsection{Resource states}
  Similarly to the case of the algebraic $\lambda$-calculus, we define resource closures, resource environments and resource states in a mutually recursive fashion.
  
  \begin{description}
    \item [Resource environment] A resource environment is a total function from the set of variables $\mathcal{V}$ to resource closures. $e_0$ is the empty environment mapping any variable in $\mathcal{V}$ to the empty closure $1$. We use the notation $[x \mapsto c]$ to refer to the environment which maps the variable $x$ to the closure $c$ and all the other variables to the closure $1$. Given two environments $e'$ and $e''$, we define their pointwise concatenation $e'e''$ such that for all variables $x$, $e'e''(x)=e'(x)e''(x)$.
      
    \item [Resource closure] A resource closure is defined as a pair $c=(T,e)$ where $T$ is a simple poly-term and $e$ is a resource environment. A resource closure is said to be elementary when its multiset $T$ is a singleton. The empty closure is $1 = (1, e_0)$. We use letters $c, c_1, \dots$ for general resource closures and $\gamma, \gamma_1, \dots$ for elementary resource closures.
    
    \item [Resource state] A resource state is a triple $(t,e,\pi)$ where $(t,e)$ is an elementary resource closure and where $\pi$ is a stack of resource closures. We denote the set of resource states $\mathcal{S}(\Delta)$.
  \end{description}
  
%  Similarly to the case of the algebraic states, the intuition behind resource states can be made explicit by defining the function $\mathrm{T}_{\mathrm{D}} : \mathcal{S}(\Delta) \rightarrow \mathbb{S} \langle \Delta \rangle$ which given a resource state returns its unique associated sum of resource terms. 
%  
%  We define $\mathrm{T}_{\mathrm{D}}$ on any resource closure $(T,e)$ then extend it to any resource state $(t,e,(c_1, \dots, c_n))$ with $n$ being a non-negative integer:
%\begin{align*}
%\label{eq:TD-def}
%  \mathrm{T}_{\mathrm{D}}(T,e) &= \partial_{x_1,\dots,x_n}(T,\mathrm{T}_{\mathrm{D}}(e(x_1)),\dots,\mathrm{T}_{\mathrm{D}}(e(x_n)))\\
%  \mathrm{T}_{\mathrm{D}}(t,e,(c_1,\dots,c_n)) &= \langle \dots \langle \mathrm{T}_{\mathrm{D}}(t,e) \rangle \mathrm{T}_{\mathrm{D}}(c_1) \dots \rangle \mathrm{T}_{\mathrm{D}}(c_n)
%\end{align*}
%
%where $\partial$ has been extended to $\Delta^{(!)} \times \mathbb{S} \langle \Delta^{(!)} \rangle$. 

%  We also define $\mathrm{T}_{\mathrm{D}} : \mathcal{S}(\Delta) \rightarrow \Delta$ similarly to T which acts on resource terms.
%For any resource closure $(T,e)$ and any resource state $(t,e,(c_1, \dots, c_n))$:
%\begin{align}
%\label{eq:TD-def}
%  \mathrm{T}_{\mathrm{D}}(T,e) &= \partial_{x_1,\dots,x_n}(T,\mathrm{T}_{\mathrm{D}}(e(x_1)),\dots,\mathrm{T}_{\mathrm{D}}(e(x_n)))\\
%  \mathrm{T}_{\mathrm{D}}(t,e,(c_1,\dots,c_n)) &= \langle \dots \langle \mathrm{T}_{\mathrm{D}}(t,e) \rangle \mathrm{T}_{\mathrm{D}}(c_1) \dots \rangle \mathrm{T}_{\mathrm{D}}(c_n)
%\end{align}
      
\section{Quantitative Krivine machine and Taylor expansion}

\subsection{Quantitative Krivine machine}

In turn we define our quantitative Krivine machine (\textbf{qKAM}) $K$\footnote{This machine has been implemented and is available online at \url{http://allioux.iiens.net/taylor/}.} which draws its inspiration from the one described in~\cite{BKT}. This definition is the main contribution of this paper. It is important to note that for the sake of convenience we will only consider closed algebraic terms which reduce to the constant $c_0$. From now on, we therefore enrich the syntax of the resource $\lambda$-calculus with this same constant $c_0$.

The following machine computes a coefficient associated with an algebraic state and a resource state. We remind that an algebraic state corresponds to a unique $\lambda$-term and a resource state corresponds to a unique sum of resource terms. Therefore, in the case of algebraic terms whose sums correspond to probability distributions, this coefficient will be the sum of the probabilities that each resource term in the sum describe a resource usage of the reduction of the algebraic term to $c_0$.

\begin{definition}(Quantitative Krivine machine)
  \label{def:qkam}
  The quantitative Krivine machine is defined as a matrix $K \in \mathbb{S}^{\mathcal{S}(\Lambda_{\mathbb{S}})\times \mathcal{S}(\Delta)}$. It is defined by induction on the pair $(\mathrm{size}(t,e,\pi), \mathrm{size}(M,E,\Pi))$\footnote{The size of a term is its number of symbols.} lexicographically ordered. $K(M,E,\Pi)_{(t,e,\pi)}$ denotes the coefficient in $K$ associated with the pair $((M,E,\Pi),(t,e,\pi))$.
  
  \begin{itemize}
  \item $K(c_0,E,\emptyset)_{(c_0,e_0,\emptyset)}=1$,
  \item $K(x,E,\Pi)_{(x,e,\pi)}=K(E(x),\Pi)_{(e(x),\pi)}$ if $x \in \mathrm{Dom}(E)$ and $e$ is such that $\forall y \neq x, e(y)=1$,
  \item $K(\lambda x.M,E,\Gamma::\Pi)_{(\lambda x.u,e,c::\pi)}=K(M,E_{x \mapsto \Gamma},\Pi)_{(u,e_{x \mapsto c},\pi)}$ if $e(x)=1$ and where w.l.o.g $x \notin Dom(E)$,
  \item $K((M)N,E,\Pi)_{(\langle t \rangle T,e,\pi)} = \sum_{\substack{(e',e'')\\e'e''=e}} K(M,E,(N,E)::\Pi)_{(t,e',(T,e'')::\pi)}$,
  \end{itemize}
    
    The major difference with the case of the ordinary $\lambda$-calculus appears in the following two cases: 
    
  \begin{itemize}
  \item $K(\alpha M,E,\Pi)_{(t,e,\pi)}=\alpha K(M,E,\Pi)_{(t,e,\pi)}$,
  \item $K(M+N,E,\Pi)_{(t,e,\pi)} = K(M,E,\Pi)_{(t,e,\pi)} + K(N,E,\Pi)_{(t,e,\pi)}$,
  \item Otherwise $K(M,E,\Pi)_{(t,e,\pi)}=0$.   
  \end{itemize}
  
\end{definition}

As we do not want to deal with states directly, we hide them by defining $\hat{K}$ which takes terms instead of states.
\begin{definition}
For any algebraic term $M$ and any resource term $t$,
  \begin{equation}
    \hat{K}(M)_t=K(M,\emptyset,\emptyset)_{(t,e_0,\emptyset)}
  \end{equation}
\end{definition}

This machine is defined for all semirings and in the particular case of $\mathbb{Q}^+$ computes a coefficient we shall characterize in Theorem~\ref{th:main-result}.

We shall give some examples of execution. Let $\Delta = \lambda x.(x)x, I=\lambda x.x, T=\lambda xy.x$ and $F = \lambda xy.y$. Consider the two examples $(\Delta) I c_0$ and $(\Delta) (pI+qF)c_0$, where $p,q \in \mathbb{S}$.
  
  \begin{equation*}
  \hat{K}((\Delta) I c_0) =
  \begin{cases}
    \langle \langle \lambda x.\langle x \rangle x \rangle (\lambda x.x)^2 \rangle c_0 \mapsto 1\\
    \_ \mapsto 0
  \end{cases}
  \end{equation*}
    
  Table \ref{tab:qkam-example} exposes the succession of states taken by the machine which are associated with a nonzero coefficient during the execution of this example. In fact, in this very case all the states have the coefficient 1 in $K$. We shall detail the transition from the $4^{th}$ to the $5^{th}$ state as this is the only one which involves a sum with several summands even though only one of these summands is nonzero.
  
  Let $\mathbf{S}_1$ be the algebraic state $((x)x, \{x \mapsto (\lambda x.x,\emptyset)\},[(c_0, \emptyset)])$ and let $\mathbf{S}_2$ be the algebraic state $(x,\{x \mapsto (\lambda x.x,\emptyset)\},[(x,\{x \mapsto (\lambda x.x,\emptyset)\});(c_0, \emptyset)])$.
  
  Then the transition from the $4^{th}$ to the $5^{th}$ state in Table~\ref{tab:qkam-example} given by Definition~\ref{def:qkam} is:
  
  \begin{align*}
  \begin{split}
    K(\mathbf{S}_1)_{(\langle x \rangle x,\{x \mapsto ((\lambda x.x)^2, e_0)\},[(c_0, e_0)])} 
    &= K(\mathbf{S}_2)_{(x, \{ x \mapsto (\lambda x.x, e_0) \}, [(x, \{ x \mapsto (\lambda x.x, e_0) \});(c_0, e_0)])}\\
    &\quad + K(\mathbf{S}_2)_{(x, \{ x \mapsto 1 \}, [(x, \{ x \mapsto ((\lambda x.x)^2, e_0) \});(c_0, e_0)])}\\
    &\quad + K(\mathbf{S}_2)_{(x, \{ x \mapsto ((\lambda x.x)^2, e_0) \}, [(x, \{ x \mapsto 1 \});(c_0, e_0)])}
  \end{split}
  \end{align*}
  
  But both $K(\mathbf{S}_2)_{(x, \{ x \mapsto 1 \}, [(x, \{ x \mapsto ((\lambda x.x)^2, e_0) \});(c_0, e_0)])}$ and $K(\mathbf{S}_2)_{(x, \{ x \mapsto ((\lambda x.x)^2, e_0) \}, [(x, \{ x \mapsto 1 \});(c_0, e_0)])}$ are equal to 0 according to Definition~\ref{def:qkam}.
  
  That is why we only show the pair of states $(\mathbf{S}_2, (x, \{ x \mapsto (\lambda x.x, e_0) \}, [(x, \{ x \mapsto (\lambda x.x, e_0) \});(c_0, e_0)]))$ in Table~\ref{tab:qkam-example}.
    
%\begin{landscape}
 \begin{center}
 \begin{table*}
  \resizebox{\columnwidth}{!}{
 	\begin{tabular}{|c|c|c||c|c|c|}
 	\hline
 	\multicolumn{3}{|c||}{Algebraic state} & \multicolumn{3}{|c|}{Resource state}\\
 	\hline
 	Term & Env. & Stack & Term & Env. & Stack\\
 	\hline
 	$((\lambda x.(x)x) \lambda x.x) c_0$ & $\emptyset$ & $[]$  & $\langle \langle \lambda x.\langle x \rangle x \rangle (\lambda x.x)^2 \rangle c_0$ & $e_0$ & $[]$\\
 	\hline
 	$(\lambda x.(x)x) \lambda x.x$ & $\emptyset$ & $[(c_0, \emptyset)]$  & $\langle \lambda x.\langle x \rangle x \rangle (\lambda x.x)^2$ & $e_0$ & $[(c_0, e_0)]$\\
 	\hline
 	$\lambda x.(x)x$ & $\emptyset$ & $[(\lambda x.x,\emptyset);(c_0, \emptyset)]$  & $\lambda x.\langle x \rangle x$ & $e_0$ & $[((\lambda x.x)^2, e_0);(c_0, e_0)]$\\
 	\hline
 	$(x)x$ & $\{x \mapsto (\lambda x.x,\emptyset)\}$ & $[(c_0, \emptyset)]$  & $\langle x \rangle x$ & $\{x \mapsto ((\lambda x.x)^2, e_0)\}$ & $[(c_0, e_0)]$\\
 	\hline
 	$x$ & $\{x \mapsto (\lambda x.x,\emptyset)\}$ & $[(x,\{x \mapsto (\lambda x.x,\emptyset)\});(c_0, \emptyset)]$  & $x$ & $\{ x \mapsto (\lambda x.x, e_0) \}$ & $[(x, \{ x \mapsto (\lambda x.x, e_0) \});(c_0, e_0)]$\\
 	\hline
 	$\lambda x.x$ & $\emptyset$ & $[(x,\{x \mapsto (\lambda x.x,\emptyset)\});(c_0, \emptyset)]$  & $\lambda x.x$ & $e_0$ & $[(x, \{ x \mapsto (\lambda x.x, e_0) \});(c_0, e_0)]$\\
 	\hline
 	$x$ & $\{x \mapsto (x,\{x \mapsto (\lambda x.x,\emptyset)\})\}$ & $[(c_0, \emptyset)]$  & $x$ & $\{ x \mapsto (x, \{ x \mapsto (\lambda x.x, e_0) \})\}$ & $[(c_0,e_0)]$\\
 	\hline
 	$x$ & $\{x \mapsto (\lambda x.x,\emptyset)\}$ & $[(c_0, \emptyset)]$  & $x$ & $\{ x \mapsto (\lambda x.x, e_0) \}$ & $[(c_0,e_0)]$\\
 	\hline
 	$\lambda x.x$ & $\emptyset$ & $[(c_0, \emptyset)]$  & $\lambda x.x$ & $e_0$ & $[(c_0,e_0)]$\\
 	\hline
 	$x$ & $\{ x \mapsto (c_0, \emptyset) \}$ & $[]$  & $x$ & $\{ x \mapsto (c_0,e_0) \}$ & $[]$\\
 	\hline
 	$c_0$ & $\emptyset$ & $[]$  & $c_0$ & $e_0$ & $[]$\\
 	\hline
  	\end{tabular}
  	}
  	\caption{Breakdown of the execution of the Krivine machine}
    \label{tab:qkam-example}
  \end{table*}
 \end{center}
%\end{landscape}

We will not give the full breakdown of the execution of the machine for the next example.
\begin{equation*}
  \hat{K}((\Delta) (pI+qF)c_0)=
  \begin{cases}
    \langle \lambda x.\langle x \rangle x\rangle I^2 \rangle c_0 \mapsto p^2\\
    \langle \lambda x.\langle x \rangle 1\rangle F \rangle c_0 \mapsto q\\
    \_ \mapsto 0
  \end{cases}
\end{equation*}

There are two non-deterministic reductions of $(\Delta) (pI+qF)c_0$ which lead to $c_0$. The first one with multiplicity $p^2$ and the second one with multiplicity $q$ which correspond to the two non-deterministic choices induced by the sum $pI+qF$.
 
\subsection{Taylor expansion}

In this setting we choose to restrict $\mathbb{S}$, the semiring over which is defined our algebraic $\lambda$-calculus, to any semiring having a multiplicative inverse such as $\mathbb{Q}^+$. Taylor expanding an algebraic term then comes down to expanding its applications according to the following formula:

\begin{equation}
\label{eq:taylor-expansion}
  ((P)Q)^* = \sum_{n=0}^{\infty} \frac{1}{n!} \langle P^* \rangle Q^{*n}
\end{equation}

\noindent where $M^*$ denotes the Taylor expansion of the algebraic term $M$ and where $Q^{*n}$ is the sum of multisets of cardinality $n$ whose elements are in the support of $Q^*$ associated with a coefficient we will not detail here but which can be found in the report.

We justify the terminology ``Taylor expansion'' by pointing out that in analysis the Taylor series of an infinitely differentiable function $f$ at $0$ is $\sum_{n=0}^{\infty} \frac{1}{n!} f^{(n)}(0)x^n$. This is, indeed, quite similar to the form of the Taylor expansion of the application in the $\lambda$-calculus. See~\cite{EHR03} for more details.

This operation can alternatively be defined by means of coefficients defined inductively on algebraic and resource terms. To this effect, we recall the coefficient $m$ described in~\cite{EHR08} accounting for the intrinsic contribution of a resource term $t$ to its coefficient in the Taylor expansion of an algebraic term $M$ and we introduce the weights $w$ which account for the dependance in $M$ of this coefficient.

\begin{definition}
\label{def:multiplicity}
The multiplicity $m$ of a resource term $t$ and the weight $w$ of a resource term $t$ in an algebraic term $M$ are inductively defined as follows:

\begin{align*}
  m(x)&=1 & w(x,x)&=1\\
  m(\lambda x.t)&=m(t) & w(\lambda x.t,\lambda x.M)&=w(t,M)\\
  m(\langle t \rangle T)&=m(t) \prod_{\mathclap{t \in \text{supp}(T)}} T(t)! m(t)^{T(t)} & w(\langle t \rangle T,(M)N)&=w(t,M) \prod_{\mathclap{t \in \text{supp}(T)}}w(t,N)^{T(t)}\\
  && w(t,\alpha M)&=\alpha w(t,M)\\
  && w(t,M+N)&=w(t,M)+w(t,N)
\end{align*}

\end{definition}

The coefficient $m(t)$ corresponds to the number of permutations of variable occurrences of $t$ preserving the name of the variables and letting the term $t$ unchanged. Finally, contrary to the case of the ordinary $\lambda$-calculus, the multiplicity of $t$ in the Taylor expansion of $M$ does not only depend on $t$ but also depends on $M$. The weights $w$ account for this phenomenon and represent one of the contributions of this paper.

We shall give some examples to enlighten the reader about these coefficients.

\begin{align*}
  m(\langle \lambda x.x \rangle (\langle y \rangle z^3)^2)
  &=m(\lambda x.x)2!m(\langle y \rangle z^3)^{2}\\
  &=m(x)2(m(y)3!m(z)^3)^2\\
  &=2*(3!)^2\\
  &=2*36=72
\end{align*}

As for the weights, their use is motivated by terms of the form $M+N$ and $\alpha M$.
Otherwise, if a term $M$ is a pure $\lambda$-term and not an algebraic term then for any $t \in \Delta$, $w(t,M)$ is equal to $1$ if $t \in M^*$ and $0$ otherwise. 

Consider the following example:

\begin{align*}
  w(\langle x \rangle x^3, (x)(2x+y)+(x)(x+z))
    &=w(\langle x \rangle x^3, (x)(2x+y))+w(\langle x \rangle x^3, (x)(x+z))\\
    &=w(x, x)w(x, 2x+y)^{3}+w(x, x)w(x, x+z)^{3}\\
    &=w(x,x)(2w(x,x)+w(x,y))^{3}+w(x,x)(w(x,x)+w(x,z))^{3}\\
    &=2^3+1=9
\end{align*}

Therefore, there are 9 ways to derive $\langle x \rangle x^3$ from $(x)(2x+y)+(x)(x+z)$.
%$w(\langle x \rangle y^2z^3, (x)(y+z))=w(x,x)w(y,y+z)^{2!}w(z,y+z)^{3!}=(w(y,y)+w(y,z))^{2!}(w(z,y)+w(z,z))^{3!}=1$

Finally, the expression of the Taylor expansion can alternatively be given by the following definition:

\begin{definition}(Taylor expansion) 
Given an algebraic term $M$, its Taylor expansion is:
    \begin{equation}
      M^*=\sum_{t \in \Delta} \frac{w(t,M)}{m(t)}t
    \end{equation}
\end{definition}

It is easy to show this definition leads to an inductive definition of the Taylor expansion on the shape of algebraic terms which is compatible with Equation~\ref{eq:taylor-expansion}. This motivates our terminology.

\subsection{Connection between the qKAM and the Taylor expansion}

The following theorem, which is one of the main contributions of this paper, along with the definition of the \textbf{qKAM}, links the behaviour of the \textbf{qKAM} with the Taylor expansion of algebraic terms.

\begin{theorem}
\label{th:main-result}
  For all algebraic terms $M \in \Lambda_{\mathbb{S}}$, for all resource terms $t \in \Delta$ and provided that $\mathbb{S}$ has a multiplicative inverse, 
  \begin{equation}
    \label{eq:main-result}
    \hat{K}(M)_{t} = M^*_t \mathrm{NF}(t)_{c_0}
  \end{equation}
  
  \noindent where $M_t^*$ is the coefficient of $t$ in $M^*$ and $\mathrm{NF}(t)_{c_0}$ is the coefficient of $c_0$ in $\mathrm{NF}(t)$.
\end{theorem}

This theorem is a particular case of a more general result applying to any algebraic state and any resource state. It can be found in the report.

\subsection{Computational complexity}
At first, it seemed that Theorem~\ref{th:main-result} informed us of an efficient way to compute $K(M,\emptyset,\emptyset)$. Indeed, the Krivine machine reduces $M$ to compute a subset of its Taylor expansion whereas the equation~\eqref{eq:main-result} gave hope we could obtain the same result more efficiently as the right-hand side does not involve the reduction of $M$. Although $M^*_t$ can be computed statically by means of the coefficients $m$ and $w$, it is folklore that determining $\mathrm{NF}(t)_{c_0}$ is \textbf{NP-complete}.

\bibliographystyle{eptcs}
\bibliography{refs} 
\end{document}